\documentclass[letterpaper, 10 pt, conference]{ieeeconf}  

\IEEEoverridecommandlockouts                              
\usepackage{graphicx}      
\usepackage[table,xcdraw]{xcolor}
\usepackage{amsmath}       
\usepackage[inkscapearea=page]{svg}
\usepackage{amsfonts}      
\usepackage{algorithm}    
\usepackage{todonotes}
\usepackage{placeins}
\usepackage{siunitx}
\usepackage{textpos}
\usepackage{booktabs}
\usepackage{footmisc}
\usepackage{comment}
\usepackage{makecell}

\usepackage[isbn=false,
url=true,
doi=false,
backend=biber,
style=ieee,
bibstyle=ieee, 
citestyle=numeric-comp, 
sortcites=true, 
maxbibnames=10,
giveninits=true, 
backref=false]{biblatex}
\usepackage{tikz}
\usepackage{tablefootnote}

\usepackage{algorithm,algpseudocode}
\makeatletter
\def\plist@algorithm{Alg.\space}
\makeatother

\newcommand{\fakepar}[1]{\vspace{1mm}
\noindent\textbf{#1.}}

\usepackage{tikz}
\usepackage{textcomp}
\usepackage{lipsum}

\newcommand\copyrighttext{%
  \footnotesize \textcopyright 2025 IEEE. Personal use of this material is permitted.
  Permission from IEEE must be obtained for all other uses, in any current or future
  media, including reprinting/republishing this material for advertising or promotional
  purposes, creating new collective works, for resale or redistribution to servers or
  lists, or reuse of any copyrighted component of this work in other works.}
\newcommand\copyrightnotice{%
\begin{tikzpicture}[remember picture,overlay]
\node[anchor=south,yshift=10pt] at (current page.south) {\parbox{\dimexpr\textwidth-\fboxsep-\fboxrule\relax}{\copyrighttext}};
\end{tikzpicture}%
}

\title{\LARGE \bf
Sailing Towards Zero-Shot State Estimation using Foundation Models Combined with a UKF 
} 

\author{Tobin Holtmann\textsuperscript{*}, David Stenger\textsuperscript{*}, Andres Posada-Moreno, Friedrich Solowjow, and Sebastian Trimpe
\thanks{The authors are with the Institute for Data Science in Mechanical Engineering, RWTH Aachen University, Germany, e-mail: \{firstname.lastname\}@dsme.rwth-aachen.de}
\thanks{This work is partially funded by the Deutsche Forschungsgemeinschaft (DFG, German Research
Foundation) under Germany’s Excellence Strategy -– EXC-2023 Internet of
Production -– 390621612. The authors acknowledge the computing time provided to them at the NHR Center NHR4CES at RWTH Aachen University (project number P0022034). This is funded by the Federal Ministry of Education and Research, and the state governments participating on the basis of the resolutions of the GWK for national high performance computing at universities (\href{http://www.nhr-verein.de/unsere-partner}{www.nhr-verein.de/unsere-partner}). Friedrich Solowjow is supported by the German federal state of North Rhine-Westphalia through the KI-Starter grant.}
\thanks{\textsuperscript{*}First two authors have equal contribution to this work.}%
} 

\usepackage[hidelinks]{hyperref}
\begin{document}

\maketitle
\copyrightnotice
\thispagestyle{empty}
\pagestyle{empty}

\begin{abstract}

State estimation in control and systems engineering traditionally requires extensive manual system identification or data-collection effort. However, transformer-based foundation models in other domains have reduced data requirements by leveraging pre-trained generalist models.
Ultimately, developing zero-shot foundation models of system dynamics could drastically reduce manual deployment effort. 
While recent work shows that transformer-based end-to-end approaches can achieve zero-shot performance on unseen systems, they are limited to sensor models seen during training.
We introduce the foundation model unscented Kalman filter (FM-UKF), which combines a transformer-based model of system dynamics with analytically known sensor models via an UKF, enabling generalization across varying dynamics without retraining for new sensor configurations.
We evaluate FM-UKF on a new benchmark of container ship models with complex dynamics, demonstrating a competitive accuracy, effort, and robustness trade-off compared to classical methods with approximate system knowledge and to an end-to-end approach.
The benchmark and dataset are open sourced to further support future research in zero-shot state estimation via foundation models.

\end{abstract}

\FloatBarrier
\section{Introduction} \label{sec:introduction} 

State estimation is a core problem in systems and control engineering, with accurate estimation playing a critical role in downstream tasks such as control and fault diagnosis.
Traditionally, Kalman filters are used in control engineering, which are powerful and interpretable methods for state estimation. 
The classic Kalman filter (KF) \cite{kalman_new_1960} combines models of system dynamics and sensor measurements to estimate the system state, including state components that are not measured directly.
While the KF relies on linear systems theory, there are also extensions to nonlinear systems, e.g., the unscented Kalman filter (UKF) \cite{julier_new_1997}.

\begin{figure*}[t]
    \centering 
    \includegraphics[width=1.7\columnwidth]{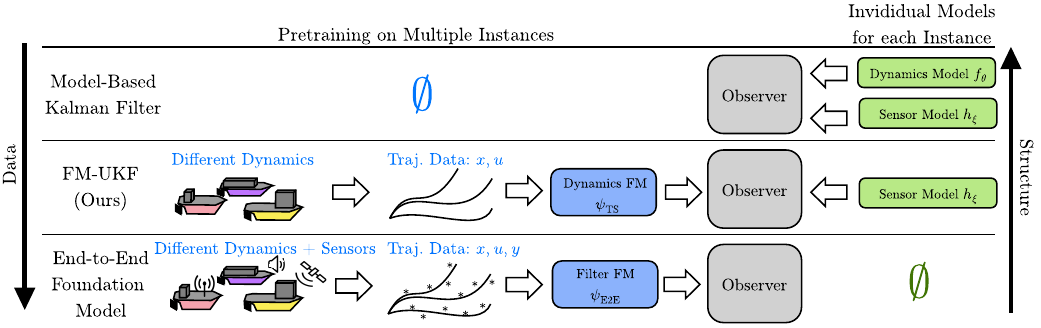}
    \caption{
    Overview of state estimation approaches. 
    Top: Classical Kalman filtering using hand-crafted system and sensor models. 
    Bottom: End-to-end transformer-based state estimation that requires extensive data collection across varying sensor configurations and dynamics. 
    Middle: The proposed FM-UKF integrates a foundation model for dynamics with analytically derived sensor models enabling robust state estimation across different sensor configurations with minimal additional effort.} 
    \label{fig:overview}
\end{figure*}

The effectiveness of these filters requires 
reliable system and sensor models. 
Deriving these models from first principles can be tedious and demands a deep understanding of system dynamics. 
Even when employing learned models, such as neural networks \cite{liu2024neural}, significant effort is required in collecting high-quality training data and tailoring the learning pipeline for each specific system (see Fig.~\ref{fig:overview}, top).

In this work, we explore the potential of the emerging foundation model paradigm \cite{bommasani_opportunities_2022} to mitigate these challenges.
Rather than training a new model for every task, a single, powerful model is trained on a wide range of related tasks, allowing for rapid adaptation. 
These models can adapt to a new instance of an unseen task with minimal or even no additional training—an approach referred to as zero-shot inference or in-context learning (ICL).
This paradigm, which originated in natural language processing with models like Google's BERT \cite{devlin_bert_2019} and OpenAI's GPT-2 \cite{radford_language_nodate}, is now gaining popularity in dynamical systems \cite{ziegler2024foundation, seifner2024foundational, song2024fmint}. 
In particular, Busetto et al. \cite{Busetto24__In_Context_learning} demonstrated an end-to-end transformer approach that maps sequences of controls and measurements directly to state estimates for a two-state system and a single sensor configuration (see Fig.~\ref{fig:overview}, bottom).
However, applying this end-to-end method to different sensor configurations would require data collection and retraining. 
Specifically, if an out-of-distribution sensor configuration is encountered, additional retraining of fine-tuning is needed.

In many practical settings sensor models are readily available from sensing principles such as obtaining position, orientation, and Allan variance from an inertial measurement unit (IMU) thus eliminating the need to learn them from scratch. 
Thus, we propose the \emph{foundation model unscented Kalman filter} (FM-UKF), which combines the foundation model paradigm to generalize over different dynamics and a modified UKF that utilizes analytically known sensor models (see Fig. \ref{fig:overview}, middle). This way, we preserve th KF's interpretability and generalization capabilities. 
Unlike the end-to-end approach \cite{Busetto24__In_Context_learning}, FM-UKF can incorporate new sensor models without fine tuning and can be trained with a single sensor configuration.
We demonstrate FM-UKF for a class of container ship models in simulation, comparing it to the end-to-end approach and classical baselines.

In summary, our contributions are:
(I) We extend results of \cite{Busetto24__In_Context_learning} showing that foundation models are effective for the state estimation problem and generalize across different complex dynamics and sensors.  
(II) We introduce FM-UKF, a novel method that combines a transformer-based dynamics model with given sensor models using a modified UKF.  
(III) We provide an open-source synthetic benchmark dataset for container ship zero-shot state estimation with 400k trajectories and 1000 different parametrization. %
(IV) We %
compare our approach against both an end-to-end transformer-based baseline \cite{Busetto24__In_Context_learning} and classical state estimation methods, %
highlighting the advantages of our proposed framework.

\section{Background \& Related Work} 

In this section, we introduce the key concepts and works related to foundation models for state estimation. 
We first introduce the general problem setting and the combination of KFs with learning approaches.
Then we present the idea of end-to-end state estimation and the general paradigm of time series foundation models.

\subsection{General Problem Setting}

We consider discrete-time non-linear dynamics:
\begin{align}
    x_{k+1} &= f_\theta(x_k,u_k) + \epsilon_k, \\
    y_k &= h_\xi (x_k, u_k) + \eta_k, \label{eq:sensor_definition}
\end{align}
where $x_k \in \mathbb{R}^{d_x}$ is the state, $u_k \in \mathbb{R}^{d_u}$ the control input, $\epsilon_k \in \mathbb{R}^{d_x}$ independent and identically distributed (iid) process noise.
We can measure $y_k \in \mathbb{R}^{d_y}$, which is corrupted by iid measurement noise $\eta_k$, and $u_k$, which is generated by control law $u_k=\pi(y_k)$.
The dynamics are parametrized via $\theta \in \Theta$ and the different observation possibilities via $\xi \in \Xi \subset \mathbb{N}$, which is a finite set of different sensor configurations.

The general problem that we consider is %
reconstructing the state from a sequence of observations and inputs:  
\begin{equation} \label{eq:E2E}
    \hat{x}_k = \psi(y_{0:k},u_{0:k}).
\end{equation}

\subsection{Learning-Based Kalman Filters for Single Systems}

The combination of KFs with machine learning models such as neural networks have a rich history to reduce the manual modeling effort as surveyed in \cite{BAI2023state}. For example, dynamics or observation models \cite{liu2024neural}, noise models \cite{gao2020rlakf} or the Kalman gain \cite{revach2022kalmannet} can be learned. Although preliminary work on transfer learning has been reported, e.g. \cite{cui2022combined}, most of these approaches only consider a single dynamical system $\theta^{\prime}$ and observation model $\xi^{\prime}$ resulting in expensive data collection for each unseen system $\theta^{*} \neq \theta^{\prime}$ and sensor model $\xi^{*} \neq \xi^{\prime}$ (see Fig. \ref{fig:overview}, top).  

\subsection{In-Context Learning for End-2-End State Estimation}

Busetto et al. \cite{Busetto24__In_Context_learning} explored the foundation model paradigm for state estimation to overcome the extensive system identification and data collection efforts. 
They propose a transformer, $\psi_\mathrm{E2E}$, that performs an end-to-end sequence-to-sequence mapping from previous inputs $u_{0:k}$ and measurements $y_{0:k}$ to the current state estimate $\hat{x}_k$:
\begin{equation}
        \hat{x}_k =\psi_{\mathrm{E2E}}(y_{0:k},u_{0:k})
\end{equation}
for all $\theta \in \Theta$ and $\xi \in \Xi$. 
Zero-shot state estimation is achieved through ICL, a paradigm common in natural language processing, where a new task is learned from a few examples during runtime without updating the model's parameters \cite{dong24__ICL_survey}. 
Here, $u_{0:k}$ and $y_{0:k}$ provide the context from which the transformer implicitly learns the previously unseen transition dynamics $f_\theta$ and sensor model $h_\xi$. 
To successfully do this within only a few time-steps, the transformer needs a foundation of relevant dynamical knowledge, instilled via supervised learning on trajectory data 
\begin{equation}
    \tau_{\theta,\xi,\pi} = \left(u_k,x_k,y_k \mid f_\theta,h_\xi,\pi\right)_{k=0}^L
\end{equation}
sampled from multiple related systems $\theta \in \Theta_{\text{train}} \subset \Theta$ and sensor models $\xi \in \Xi_\mathrm{train} \subset \Xi$.

In practice, applying such an end-to-end approach to a new system with an unseen sensor $\xi' \notin \Xi_\mathrm{train}$ requires retraining or fine-tuning on an expanded dataset (see Fig. \ref{fig:overview}, middle).

This method was validated on a synthetic dataset comprising 1.6 million instances of an evaporative process with a two-dimensional state space, using a simple linear sensor in which one of the two features is observed as noise.

From a theoretical point of view, Goel et al. \cite{goel_can_2024} prove under idealized conditions that there exists a stable weight configuration for a transformer whose performance matches that of a KF for linear time-invariant systems.

\subsection{Time Series Foundation Models}

Time series foundation models are large, expressive often transformer based  models that learn diverse dynamics through extensive autoregressive pretraining. 
Their key idea is to capture a wide range of mechanisms so that, when applied to downstream tasks, little to no additional data is required. 
For instance, these models are frequently employed for long-horizon forecasting, where $h$ denotes the horizon:
\begin{equation}
    \hat{x}_{k+1:k+h} = \psi_{\mathrm{TS}}(x_{0:k},u_{0:k}). \label{eq:dynfm}
\end{equation}
In forecasting tasks, these models use the context provided in each prediction to infer the underlying system dynamics via ICL and generate plausible long-horizon forecasts based on what they learned during their pre-training. 
Recent approaches, including architectures such as TimesFM \cite{das_decoder-only_2024}, Timer \cite{liu2024timer}, and MOIRAI \cite{WooLKXSS2024moirai}, focus on robust projection mechanisms and embeddings to enable scalable autoregressive training.
In comparison with traditional learning-based models, time series foundation models aim towards better generalization (models that require little or no additional data for adaptation to new tasks) thus providing more robust predictions and often improved performance \cite{bommasani2021opportunities}.

\section{Problem Formulation: Combining Dynamics Foundation Models with Kalman Filters}

In this work, we assume that the sensor model $ h_\xi(\cdot) $ is known \eqref{eq:sensor_definition}. 
This assumption enables us to use established analytical models for sensor behavior, eliminating the need for extensive calibration or learning from scratch. 
Furthermore, these models are widely applicable across domains and can be easily selected for different hardware configurations.
Our goal is to combine a dynamics foundation model $\psi_{\mathrm{TS}}$ \eqref{eq:dynfm} (with horizon $h=1$) with a KF, obtaining a hybrid estimator:
\begin{equation}
    \hat{x}_k = \psi_{\mathrm{FMUKF}}(y_{0:k},u_{0:k}; h_{\xi},\psi_{\mathrm{TS}}).
\end{equation}
In contrast to end-to-end approaches that require extensive data collection for every sensor configuration, our method only needs trajectory data
\begin{equation}
	\tau_{\theta,\pi} = \left(u_k, x_k \mid f_\theta, \pi\right)_{k=0}^{L}
\end{equation}
from multiple related systems $\theta \in \Theta_{\text{train}} \subset \Theta$. 
In practice, this trajectory data can be obtained from simulations or identified from noisy measurements using established \hbox{methods \cite{liu2024neural}}.

Crucially, by using a dynamics foundation model that generalizes across different systems, we do not need to retrain the dynamics model for each new sensor configuration and system parametrization. 
Instead, trajectory data $\tau_{\theta,\pi}$ can be obtained from systems with similar parametrization in simulation or real systems. Additionally, the known sensor model $h_\xi(\cdot)$ is integrated within the KF framework, enabling robust state estimation with minimal additional adaptation effort (see Fig. \ref{fig:overview}, bottom).



\section{Method: Foundation Model Unscented Kalman Filter (FM-UKF)} \label{sec:background and method}

In this section, we propose our foundation model unscented Kalman filter (FM-UKF)\footnote{Code and Data available at \href{https://github.com/Data-Science-in-Mechanical-Engineering/fm-ukf}{github.com/Data-Science-in-Mechanical-Engineering/fm-ukf}}. 
First, we describe how we integrate a foundation-model-based dynamics function into the KF.
Afterwards, we detail the dynamics foundation model architecture and key implementation details. 

\subsection{Integrating FM-Dynamics into UKF}
\begin{figure}[t]
    \centering
\includegraphics[width = 1\columnwidth]{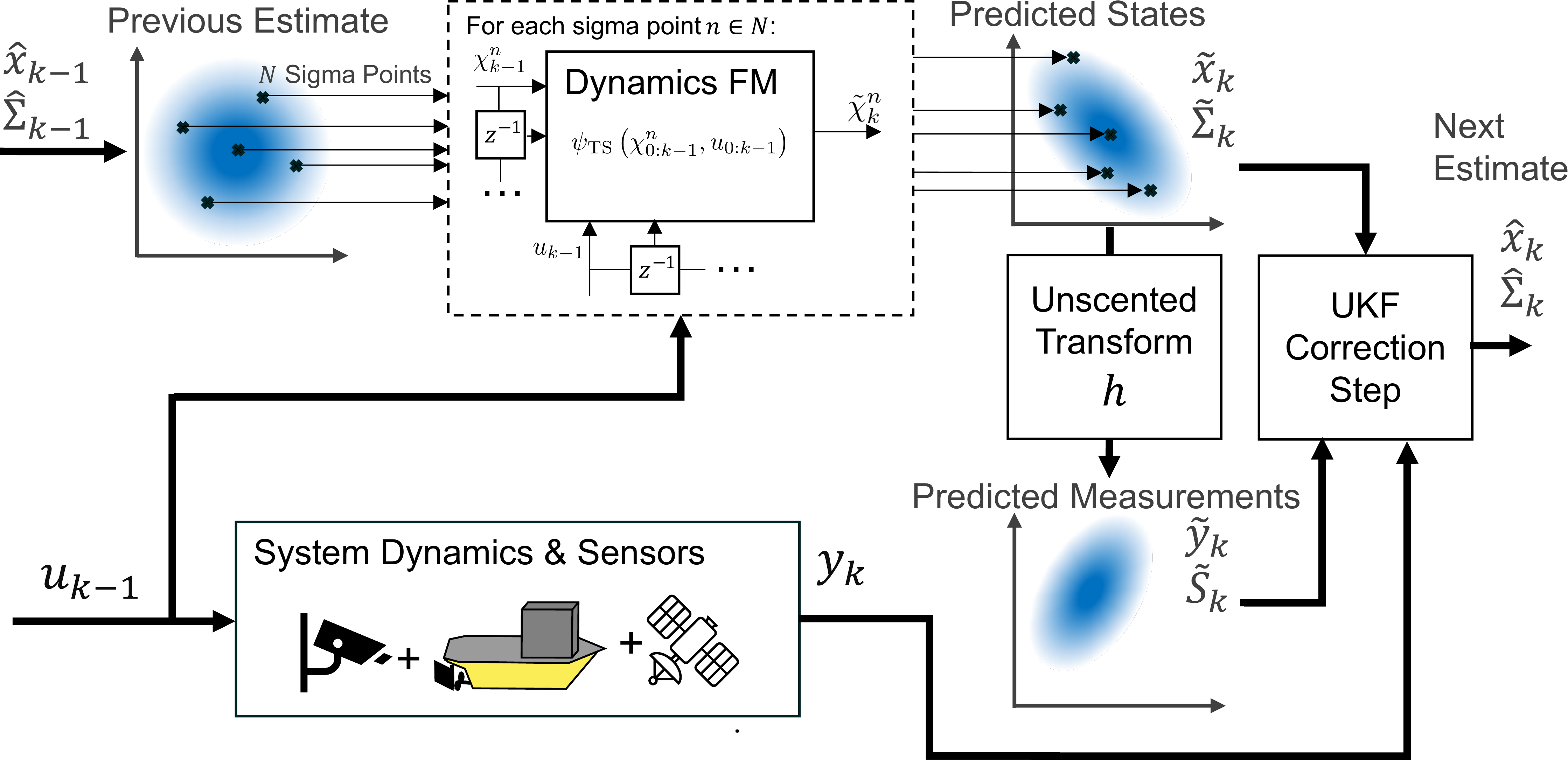}
    \caption{FM-UKF: Combining a dynamics foundation model and an UKF. A one-step time delay is denoted by $z^{-1}$ } \label{fig:methods}
\end{figure}

We propose the integration of a FM-dynamics model into an unscented Kalman filter (UKF) \cite{haykin_unscented_2001}, as shown in Fig. \ref{fig:methods}. For the FM-dynamics, we use a decoder-only transformer similar to \cite{Busetto24__In_Context_learning}. Specifically we use key architecture innovations from TimesFM \cite{das_decoder-only_2024} that is well proven for time-forecasting tasks as a basis and adjust the loss function, feature transformation and output patching to our application (see Sec. \ref{sec:architecture}, \ref{sec:training}).
We chose the UKF \cite{julier_new_1997} among other non-linear KF variants because it is computationally efficient and easy to implement: In contrast to the extended KF, the UKF does not rely on the linearization of the potentially non-smooth FM-dynamics, and unlike the particle filter \cite{moral1996nonlinear}, the UKF only requires relatively few forward passes through the FM-dynamics, reducing the computational load.

The UKF employs the unscented Transformation (UT) \cite{uhlmann_dynamic_1995-1} to propagate the uncertainties through non-linear dynamics and sensor models. First, the previous state estimate distribution $\mathcal{N}(\hat{x}_{k-1}, \hat{\Sigma}_{k-1})$ is represented as an ensemble $\mathcal{X}_{k-1}=\{\chi_{k-1}^n\ \in\mathbb{R}^{d_x}\}^N_{n=0}$  (known as sigma points) and weights $\{W^n_\mu\ \in \mathbb{R}\}^N_{n=0}$  and  $\{W^n_\Sigma\ \in \mathbb{R}\}^N_{n=0}$. Then each sigma point is propagated through the dynamics model
\begin{equation} \label{eq:UKF-classic-sigma-forward}
\ \tilde{\chi }_k^{n}\leftarrow f\left( \chi_{k-1} ^{n} ,{u}_{k-1}\right) \ \forall \chi_{k-1} ^{n} \in \mathcal{X}_{k-1}.
\end{equation}
The parameters of the gaussian approximation $\mathcal{N}(\tilde{x}_k, \tilde{\Sigma}_{k})$ of the predicted estimate distribution is obtained with
\begin{equation}\label{eq:UKF-Sigma-Predict}
\tilde{x}_k = \sum_{n=0}^{N} W_\mu^n \tilde{\chi}_k^n  \ \ \tilde{\Sigma} = \sum_{n=0}^{N} W_\Sigma^n \left( \tilde{\chi}_k^n- \tilde{x}_k \right)\left( \tilde{\chi}_k^n - \tilde{x}_k \right)^\top .
\end{equation}
The predicted measurement distribution is obtained with another UT. 

Unlike a classical UKF, which uses a user-specified function~\(f\) for its prediction step, we use a dynamics foundation model. 
The FM-dynamics use a history of previous states and control inputs $x_0,u_0,...,x_k,u_k$ to predict the next state $ {x}_{k+1}$, inferring the dynamics from the context and using them to predict the next state.
Consequently in the prediction step, instead of propagating each sigma point $ \chi ^{n}$ \ through $ f$ just at the current time-step, the sigma points sets are stored for every timestep, and used as the history context for the transformer. 
This modifies  \eqref{eq:UKF-classic-sigma-forward} in the prediction step to:
\begin{equation}
\tilde{\chi }_{k}^{n}\leftarrow \psi_{\mathrm{TS}} \left( \chi _{0:k-1}^{n} ,{u}_{0:k-1}\right)
\end{equation}

This approach assumes that the sequence of each $n$-th sigma point trajectory forms a coherent and sufficient context for the transformer to infer the dynamics.  The weights and sigma points are calculated according to the "base set" as recommended by S. Bitzer \cite{bitzer_sbitzerukf-exposed_2024}.
\subsection{Dynamics Foundation Model Architecture} \label{sec:architecture}

We base our approach on the TimesFM architecture~\cite{das_decoder-only_2024}, a decoder-only transformer. 
This architecture has been shown to be effective for time-series forecasting---a task closely related to dynamics modeling. 
Its core idea is to learn an autoregressive mapping from past inputs to future states. 
In fact, without patching the raw sequence, the architecture is almost identical to that used by Busetto et al. in their end-to-end state estimator~\cite{Busetto24__In_Context_learning}. 

The selected model architecture is composed of three key sequential modules. 
(I) First, we embed the input sequence into a series of tokens.  
(II) Second, we process these tokens through a series of attention layers. 
(III) Third, we project the obtained tokens into the output sequence, reversing the projection and patching from the first step. 
The models are trained following the typical transformer training paradigm, exploiting causal masking to process the entire input sequence in one forward pass, which enables independent predictions for every time-step in the output sequence and allows for the loss computation.

The input sequence $a_{0:L-1}$ where ${a}_{k} = \text{concat}({x}_{k}, {u}_{k})$ is embedded into a series of tokens in a three-step process. 
First, the input time series of length $L$ into non-overlapping segments, or `patches', each containing $p$ consecutive time steps. 
In our implementation, we utilize a patch size of $p=4$, transforming the sequence length from $L$ to $L/p$ while increasing the feature dimension of each token proportionally.
Next, these patches are projected through a residual block into a latent dimension, and finally, a learnable positional encoding is added to incorporate temporal information. 
This embedding process was introduced in TimesFM~\cite{das_decoder-only_2024} and ensures that both local and global temporal features are adequately captured.

After obtaining the tokenized representation, we process it through a stack of $N$ multi-headed self-attention layers~\cite{vaswani2017attention}. 
Causal masks within these attention modules ensure that each token’s representation is computed using only information from the current and preceding time steps.
This mechanism is critical for respecting causality during inference.

Finally, the latent tokens are projected back from the latent space to the desired output dimension and then depatched to recover the output sequence $b_{0:L-1}$ where ${b}_{k} = {x}_{k+1}$. 
This final projection utilizes a residual block structurally similar to the input embedding's projection layer, mapping the processed tokens back towards the target dimension. 
Subsequently, the `depatching' operation reverses the initial patching by reorganizing the features within each token back into their corresponding time-series components, obtaining the final sequence prediction.

\subsection{Training} \label{sec:training}

Training the dynamics foundation model requires several techniques to ensure stable convergence and accurate dynamics learning. 
In contrast to TimesFM, we adopted a modified input/output scheme called Masked Long-Horizon Prediction (MLHP), a Huber loss function, and a modified feature transformation and scaling scheme.

\fakepar{Masked Long Horizon Prediction} 
A key component was the adoption of MLHP, which forces the model to predict future states based on a limited initial state-action context and subsequent control inputs only, driving it to internalize system dynamics. With MLHP input sequences $a_{0:L-1}$ and target output sequence $b_{0:L-1}$ are structured as:
\begin{equation}
\begin{aligned}
a_{k} & =\begin{cases}
\text{concat} ({x}_{k} +\epsilon _{x} ,{u}_{k} +\epsilon _{u} ) & k< L_{\mathrm{context}}\\
\text{concat} ({0} ,{u}_{k} +\epsilon _{u} ) & k \geq L_{\mathrm{context}}
\end{cases}\\
b_{k} & ={x}_{k+1}.
\end{aligned}
\label{eq:MLHP} 
\end{equation}
Here, $L_{\mathrm{context}}$ defines the context length, and $\epsilon_x$, $\epsilon_u$ are small Gaussian noise terms to reduce overfitting. After the context window, states are masked (set to zero), forcing the model to rely on its learned dynamics and control inputs to predict ${x}_{k+1}$. The context length is randomly varied between 64 and 84 during training to prevent overfitting to a fixed context size.

\fakepar{Loss Function} 
While standard Mean Squared Error (MSE) loss proved sensitive to outliers and gradient explosions, we found a normalized Huber loss \cite{huber_robust_1964} more effective. 
The Huber loss combines quadratic loss for small errors ($|x| < \delta$) with linear loss for large (sporadic) errors ($|x| \geq \delta$), defined as:
\begin{equation}
\mathcal{L}_{\delta }( x) =\begin{cases}
x^{2} /2 & |x|< \delta \\
\delta ( |x|-\delta /2) & |x|\geq \delta.
\end{cases}
\end{equation}
We set $\delta=1$. 
To account for varying scales and dynamics across different state features, we normalize the prediction error for each feature $j$ by the inverse of its average absolute step-to-step change in the target trajectory. 
This focuses the loss on slow-moving features where small errors are more significant. 
The final loss function is:
\begin{equation}
\begin{aligned}
& \mathcal{L}=\frac{1}{L}\sum _{k=1}^{L-1}\sum _{j=1}^{d_x} \mathcal{L}_{\delta =1}\left(\frac{\left(\mathbf{x}_{k}^{\text{target}} -\mathbf{x}_{k}^{\text{pred}}\right)_{j}}{\frac{1}{L-1}\sum _{k'=0}^{L-2} |\mathbf{x}_{k'+1}^{\text{target}} -\mathbf{x}_{k'}^{\text{target}} |_{j}}\right)\\
\end{aligned}
\label{eq:Final-Huber-loss}
\end{equation}
This normalization provides a consistent baseline loss across features and instances, aiding evaluation.

\fakepar{Feature Transformations and Padding}
Before being processed by the model, input features are transformed for stability and consistency. 
Every feature dimension is independently normalized using standard scaling, calculating the mean and standard deviation across the entire training dataset. For rotational features, such as the heading angle $\psi$ of a container ship (see Sec. \ref{sec:Method-BenchMark}), which has a discontinuity of the at its $0^\circ/360^\circ$ boundary, we use rotation vectors, transforming it into $[\sin(\psi), \cos(\psi)]$. 
This continuous representation simplifies loss computation and improves learning stability; outputs are converted back to angles using $\arctan2$ during inference. 
Since we use input patching with patch size $p=2$, sequences require a length $L$ that is a multiple of $p$. 
Whenever $L \bmod p \neq 0$, zero vectors are appended to the beginning of the time series to achieve the correct length. 

Additionally, to ensure generalization across sequence lengths, the first $r \sim U([0, p-1])$ time steps are randomly zero-masked, following the approach in \cite{das_decoder-only_2024}.

\fakepar{Regularization Methods}
To further stabilize training and improve convergence, particularly given the model complexity and challenging dynamics, several regularization strategies were employed.
\textbf{Gradient Clipping:} The L2 norm of gradients was clipped to a maximum value of 1. 
This prevents excessively large parameter updates caused by outliers or unstable dynamics, mitigating gradient explosion \cite{zhang_why_2020}.
\textbf{Learning Rate Scheduling:} An exponential decay schedule was applied to the Adam optimizer's learning rate (initial rate 0.001). 
After a warm-up period of 60 epochs, the rate was reduced by a factor of approximately 0.774 every 50 epochs, promoting smoother convergence in later training stages.
\textbf{Gradient Accumulation Scheduling:} To simulate larger batch sizes without increasing memory requirements, gradients were accumulated over multiple mini-batches before each optimizer step. 
To balance rapid initial learning with increased stability later in training, the effective batch size was increased from 256, to 2048, and then 4096 in epochs 40 and 800 respectively.
Additionally, standard \textbf{Dropout} with a rate of 0.01 was applied within the transformer layers (as listed in Table \ref{tab:Transformer-hyperparameters}).

For large scale training it is critical and non-trivial to correctly tune the relevant hyperparameters. The most important parameters of our final model are listed in Tab. \ref{tab:Transformer-hyperparameters}.

\begin{table}[!h]
\centering
\caption{Hyper-parameters of the transformer models used in experiments}
\begin{tabular}{l|l}
Total number of parameters & 1.8m \\ \hline
Training epochs            & 1000 \\ \hline
Training time              & 12h \\ \hline
Input patch size $p$       & 2 \\ \hline

Embedding dimension & 128 \\ \hline
Number of Decoder Layers & 8 \\ \hline
Number of MHSA heads     & 32 \\ \hline
Decoder MLP Neurons      & 1024  \\ \hline
Residual Block Neurons   & 1024 \\ \hline
Dropout                  & 0.01 \\

\end{tabular}
\label{tab:Transformer-hyperparameters}
\end{table}


\section{Container Ship Benchmark for Zero-shot State Estimation}\label{sec:Method-BenchMark}

Previous work \cite{Busetto24__In_Context_learning}  on end-to-end zero-shot state estimation considered a simple system with a two-dimensional state space. To evaluate zero-shot state estimation capabilities, we developed a synthetic dataset consisting of 400k trajectories of 384 time-steps (equivalent to about 6 minutes) generated from 1000 unique parameter instances $\{ {\theta_1},...,{\theta}_{1000} \} \subset \Theta $ of a non-linear model \cite{son_coupled_1981} of a containership\footnote{Code implementation adapted from  \cite{fossen_cybergalacticmss_2024}}. These ships are maneuvered by a rudder and single propeller, modeled as first order systems, with a control input vector is defined by the target rudder angle $\delta _{c}$ and propeller/shaft velocity $n_{c}$. The state vector $\boldsymbol{x}$ is formed by these two actuator states, and eight position and velocity features (see Fig. \ref{fig:results}).

This benchmark poses three key challenges: (I) the state space is ten-dimensional, (II) the non-linear dynamics model is highly sensitive to its 70 parameters $\theta$, and (III) states are defined in both a non-inertial body-fixed reference frame and a  world frame.

\subsection{Sampling of Diverse Instances of Systems}
\label{sec:SamplingUniqueInstances}
Son et al. \cite{son_coupled_1981} published the model with an experimentally verified set of parameters that we refer to as the base parameters $\theta_{\mathrm{base}}$. New candidate instances $\theta_i$ are generated by varying eleven parameters %
defining the mass, dimensions, hydrodynamics, and inertial tensors, by uniformly sampling $\pm30\%$ of their nominal base values. A candidate is rejected if it is unstable, i.e., capsizes. Furthermore, we introduce the dissimilarity metric
\begin{equation}
    \text{dsim}^{2} (\theta _{i} ,\theta _{j} )=\mathbb{E}_{{x} ,{u}}\left[\sum _{d=1}^{d_{x}}\frac{|f_{\theta _{i}}({x} ,{u}) -f_{\theta _{j}}({x} ,{u}) |_{d}^{2}}{\mathbb{E}_{{x} ,{u}}[|f_{\theta _{\mathrm{base}}}({x} ,{u}) -{x} |_{d}^{2}]}\right]
\end{equation}
and filter out approximately the $50\% $ of remaining stable candidates with the lowest nearest neighbor dissimilarity $\min_{j}\text{dsim} (\theta_i ,\theta _{j})$.  This ensures that all instances are sufficiently unique.

\subsection{Trajectory Generation}

To generate sufficiently excited and diverse trajectories, we use pink noise as the control inputs. Pink noise has shown favorable exploratory properties in other domains, such as off-policy reinforcement learning algorithms \cite{eberhard_pink_2022}. Its properties make it particularly well-suited for generating control inputs in scenarios where smooth state transitions and effective state space coverage are necessary. In practical terms, pink noise introduces time correlation between successive control inputs, meaning that the values do not change too abruptly.  
 
\subsection{Sensor Models} \label{sec:method-sensor-models}
We consider two different sensor models $h$ for our approach. Sensor configuration $h_1$ consists of noisy measurements of all states except for the actuator states. The second, more challenging configuration $h_2$  additionally does not include the surge $u$ and sway $v$ velocities. This represents a somewhat realistic approach, since on container ships, GPS is 
available, but directly measuring absolute velocities is more challenging. 

\section{Empirical Evaluation} \label{sec:virtualDataSet}

\begin{figure*}[t]
    \centering
    \includegraphics[width = 2.0\columnwidth]{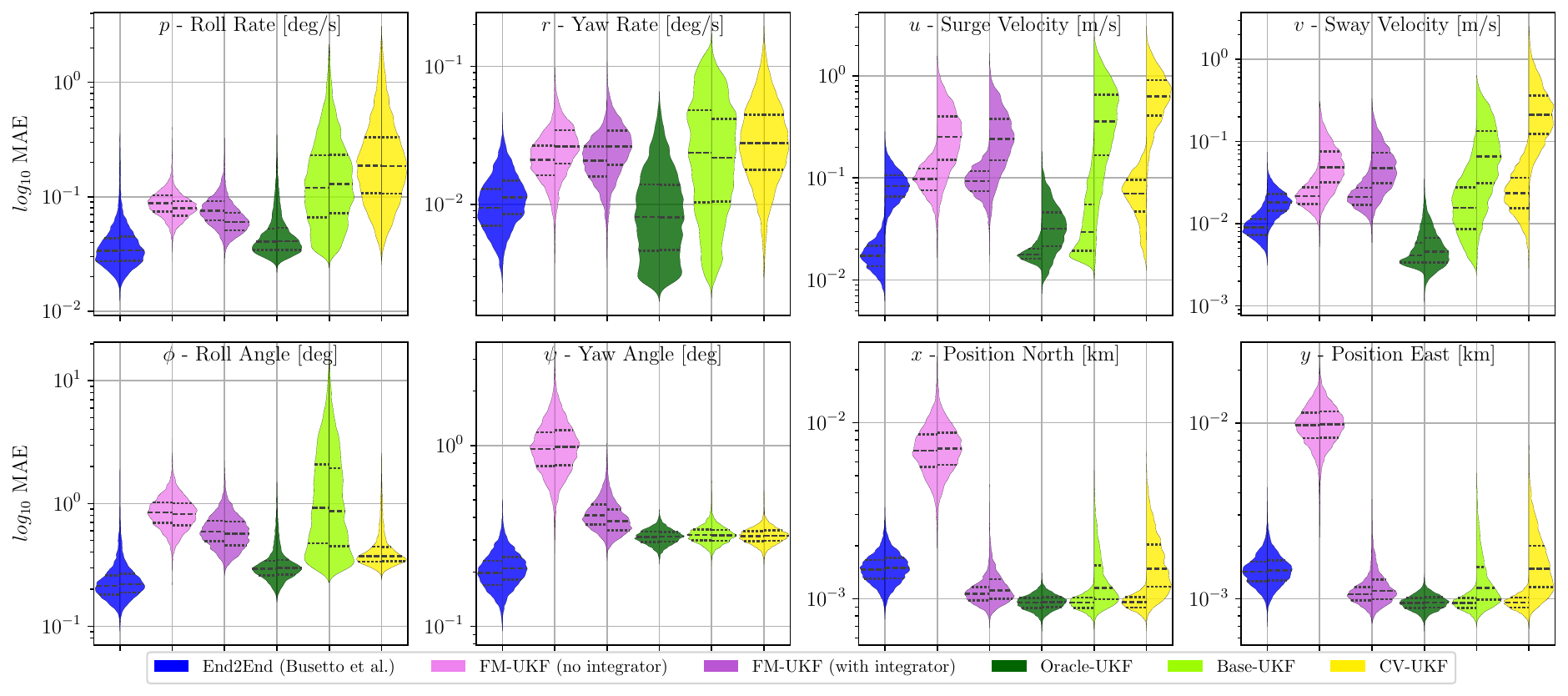}
    \vspace{-3mm}
    \caption{Distribution of mean absolute error for each zero-shot state estimation across 5000 trajectories from 100 unseen systems. 
    Left side: $h_1$ - all sensor. Right side: $h_2$ - no velocities (see Sec. \ref{sec:method-sensor-models}). The End2End approach extended from \cite{Busetto24__In_Context_learning} performs similar to the filter with perfect knowledge (Oracle-UKF) when trained on identical sensor configurations. Our approach FM-UKF (with integrator - see Sec.~\ref{sec:EvaluatedObs}) effectively generalizes over different unseen system dynamics outperforming classical baselines (Base-UKF, CV-UKF) when velocities are not measured ($h_2$). FM-UKF does not require pretraining on different sensors as End2End does.} 
    \label{fig:results}
\end{figure*}

\begin{table}[!h]
    \centering
    \caption{Median MAE and mean estimator ranks.}
    \begin{tabular}{ll|c|c|c|c} 
   &  & \makecell{End2End \\ (Busetto et al.)} & \makecell{FM-UKF \\ (integrator)} & \makecell{Base-UKF \\  \phantom{.}} & \makecell{CV-UKF \\  \phantom{.}} \\ \hline 
 $p$ & $h_1$ & $3.36 \mathrm{e}{-2}$ & $7.51 \mathrm{e}{-2}$ & $1.20 \mathrm{e}{-1}$ & $1.86 \mathrm{e}{-1}$ \\ 
 & $h_2$& $3.41 \mathrm{e}{-2}$ & $6.01 \mathrm{e}{-2}$ & $1.29 \mathrm{e}{-1}$ & $1.85 \mathrm{e}{-1}$ \\ \hline 
 $r$ & $h_1$ & $9.46 \mathrm{e}{-3}$ & $2.07 \mathrm{e}{-2}$ & $2.36 \mathrm{e}{-2}$ & $2.77 \mathrm{e}{-2}$ \\ 
 & $h_2$& $1.12 \mathrm{e}{-2}$ & $2.62 \mathrm{e}{-2}$ & $2.17 \mathrm{e}{-2}$ & $2.78 \mathrm{e}{-2}$ \\ \hline 
 $u$ & $h_1$ & $1.73 \mathrm{e}{-2}$ & $9.32 \mathrm{e}{-2}$ & $2.97 \mathrm{e}{-2}$ & $7.03 \mathrm{e}{-2}$ \\ 
 & $h_2$& $8.34 \mathrm{e}{-2}$ & $2.42 \mathrm{e}{-1}$ & $3.58 \mathrm{e}{-1}$ & $6.31 \mathrm{e}{-1}$ \\ \hline 
 $v$ & $h_1$ & $9.06 \mathrm{e}{-3}$ & $2.12 \mathrm{e}{-2}$ & $1.56 \mathrm{e}{-2}$ & $2.34 \mathrm{e}{-2}$ \\ 
 & $h_2$& $1.82 \mathrm{e}{-2}$ & $4.73 \mathrm{e}{-2}$ & $6.58 \mathrm{e}{-2}$ & $2.11 \mathrm{e}{-1}$ \\ \hline 
 $\phi$ & $h_1$ & $2.13 \mathrm{e}{-1}$ & $5.91 \mathrm{e}{-1}$ & $9.23 \mathrm{e}{-1}$ & $3.74 \mathrm{e}{-1}$ \\ 
 & $h_2$& $2.21 \mathrm{e}{-1}$ & $5.69 \mathrm{e}{-1}$ & $8.63 \mathrm{e}{-1}$ & $3.73 \mathrm{e}{-1}$ \\ \hline 
 $\psi$  & $h_1$ & $1.97 \mathrm{e}{-1}$ & $4.11 \mathrm{e}{-1}$ & $3.20 \mathrm{e}{-1}$ & $3.15 \mathrm{e}{-1}$ \\ 
 & $h_2$& $2.09 \mathrm{e}{-1}$ & $3.81 \mathrm{e}{-1}$ & $3.18 \mathrm{e}{-1}$ & $3.17 \mathrm{e}{-1}$ \\ \hline 
 $x$ & $h_1$ & $1.47 \mathrm{e}{-3}$ & $1.07 \mathrm{e}{-3}$ & $9.50 \mathrm{e}{-4}$ & $9.55 \mathrm{e}{-4}$ \\ 
 & $h_2$& $1.50 \mathrm{e}{-3}$ & $1.11 \mathrm{e}{-3}$ & $1.15 \mathrm{e}{-3}$ & $1.48 \mathrm{e}{-3}$ \\ \hline 
 $y$ & $h_1$ & $1.43 \mathrm{e}{-3}$ & $1.06 \mathrm{e}{-3}$ & $9.48 \mathrm{e}{-4}$ & $9.53 \mathrm{e}{-4}$ \\ 
 & $h_2$& $1.45 \mathrm{e}{-3}$ & $1.11 \mathrm{e}{-3}$ & $1.15 \mathrm{e}{-3}$ & $1.49 \mathrm{e}{-3}$ \\ \hline 
 rank & $h_1$ & 1.75 & 3.00 & 2.38 & 2.88 \\ 
  & $h_2$ & 1.62 & 2.25 & 2.75 & 3.38 \\
 \end{tabular} 

    \label{tab:results}
\end{table}

\label{sec:Evaluated Estimators}

\subsection{Evaluated Observers} \label{sec:EvaluatedObs}
In addition to the \textbf{FM-UKF} as defined in Sec. \ref{sec:background and method}, we include a variant in which the prediction for positions $x$, $y$ and angles $\phi$, $\psi$ is replaced with a first order midpoint integrator of the predicted velocities and angular rates. This simple kinematic model is always available and also used in all baselines (Oracle-UKF, Base-UKF, CV-UKF). 

To gauge the FM-UKF’s trade-off between system identification effort and accuracy, we apply the UKF to system models $f$ with varying levels of prior knowledge. The minimal-knowledge case is represented by the \textbf{CV-UKF}, which uses a low-effort model assuming constant velocity. The idealized case is the \textbf{Oracle-UKF}, where the correct model $\theta_i$ is always used for each instance. As a middle ground, we consider the \textbf{Base-UKF}, which applies the same (incorrect) parametrization $\theta_i \approx \theta_{\mathrm{base}}$ to all instances.

We also implemented an end-to-end \textbf{(End2End)} transformer \eqref{eq:E2E} similar to \cite{Busetto24__In_Context_learning}, but for the problem at hand, which required retraining from scratch and some adjustments. It was trained on the same dataset and transformer architecture as the FM-UKF, but with differently tuned learning rate and gradient accumulation schedule. Unlike \cite{Busetto24__In_Context_learning}, it was trained simultaneously on different sensor models, $h_1$ and $h_2$.

\subsection{Filter Hyperparameters} \label{sec:filterparams}

The UKF performance critically relies on the process and measurement noise matrices. We use the correct measurement noise parameters. For the process noise, we use the same noise parametrization for all filters and we manually tune them w.r.t. good baseline performances (Oracle-UKF, Base-UKF, CV-UKF). This process noise is not optimized for the FM-UKF, thus the presented results should be seen as a lower bound on the FM-UKF performance. In future work, the measurement noise matrices could be optimized (e.g., using Bayesian Optimization \cite{chen2018weak}), directly predicted by the dynamics FM, or estimated online with adaptive filters.

\subsection{Dataset and Evaluation Criteria  }
For training and evaluation, we randomly select 900 ships $\Theta_\mathrm{train} = \{\theta_1,\dots \theta_\mathrm{900}\}$  from our dataset (Sec. \ref{sec:Method-BenchMark}). The FM-UKF and baseline estimators are evaluated on 5k trajectories of 192 time steps from the remaining 100 ships, using $h_1$ and $h_2$ defined in Sec. \ref{sec:method-sensor-models}. As evaluation criteria, we consider the distribution of the mean absolute error (MAE) across each trajectory and feature (see Fig. \ref{fig:results}). For select estimators, we also report the median of these error distributions and the estimator’s average rank (lower is better) in terms of lowest median error per feature in decreasing order in Table \ref{tab:results}.

\subsection{Results \& Discussion}

First, we observe that the Base-UKF performs a lot worse than the Oracle-UKF. This confirms that the sampled models (see Sec. \ref{sec:SamplingUniqueInstances}) are sufficiently different to the point where the CV-UKF heuristic performs comparable to the Base-UKF. Thus, using different models in the filter for the different system instances is critical.

We observe that the End2End approach closely matches the oracle UKF performance with the exception of the highly integrated states north east and yaw-angle. This extends the finding of \cite{Busetto24__In_Context_learning}, where the approach was evaluated on a two-dimensional state space to a substantially more complex setting. However, in contrast to the FM-UKF, this approach requires retraining for each unobserved sensor configuration. 

For the highly integrated states especially position and yaw angle, the FM-UKF (no integrator) performs drastically worse than the other approaches. This affect vanishes, when the first-principle kinematics equations are used for the integration. We hypothesize that indefinite integration is difficult for transformers to learn.

In the ``all sensor ($h_1$)'' case (left violins), the FM-UKF with integrator performs comparable or worse except for the roll and yaw rates than the CV-UKF and Base-UKF baselines. This may be attributed to not-sufficiently-tuned process nose (cf. Sec. \ref{sec:filterparams}).  However, in the ``no-velocities ($h_2$)'' case where more model knowledge is required, the FM-UKF outperforms the CV-UKF and Base-UKF baselines. This highlights that the transition dynamics are learned by the foundation model and effectively integrated in the UKF. In fact, using the FM-dynamics leads to a better performance than using a constant velocity model (CV-UKF) or a correct model structure with detuned parameters (Base-UKF).

\section{Conclusion \& Outlook} \label{sec:Conclusion}
This work explored the usage of foundation models on a newly developed and open sourced container ship state estimation benchmark. First, it was shown that the end-to-end approach \cite{Busetto24__In_Context_learning} is applicable to more complex systems than in the original study \cite{Busetto24__In_Context_learning} and can generalize over different sensor configurations. Second, we proposed FM-UKF, a combination of FM-dynamics with an UKF. The FM-UKF combines machine learning and control engineering ideas to avoid retraining for each new sensor configuration while generalizing over different dynamics. We showed that FM-UKF can outperform classical baselines in specific scenarios. Future work could address a more sophisticated way of propagating the uncertainties through the transformer, for instance, by directly predicting the covariance matrices, as has been done with RNNs for single systems \cite{jung_mnemonic_2020}.

\FloatBarrier

\AtNextBibliography{\footnotesize}

\printbibliography

\addtolength{\textheight}{-6cm}

\newpage

\end{document}